\documentclass[aps,pre,twocolumn,groupedaddress,showpacs]{revtex4}

\usepackage{amsmath}
\usepackage{amssymb}
\usepackage{epsfig}

\begin{document}


\title{Asymptotic decay of pair correlations in a Yukawa fluid}


\author{P. Hopkins}
\author{A.J. Archer}
\email[]{Andrew.Archer@bristol.ac.uk}
\author{R. Evans}
\affiliation{H.H. Wills Physics Laboratory,
University of Bristol, Bristol BS8 1TL, United Kingdom}


\date{\today}

\begin{abstract}
We analyse the $r \rightarrow \infty$ asymptotic decay of the total
correlation function, $h(r)$, for a fluid composed of particles interacting via
a (point) Yukawa pair potential. Such a potential provides a simple model for
dusty plasmas. The asymptotic decay
is determined by the poles of the liquid structure factor in the complex plane.
We use the hypernetted-chain closure to the Ornstein-Zernike equation
to determine the line in the phase diagram,
well-removed from the freezing transition line, where crossover occurs
in the ultimate decay of $h(r)$, from monotonic to damped oscillatory. We show:
i) crossover takes place via the same mechanism (coalescence of imaginary poles)
as in the classical one-component plasma and in other models of Coulomb
fluids and ii) leading-order pole contributions provide an accurate description
of $h(r)$ at intermediate distances $r$ as well as at long range.
\end{abstract}


\maketitle

The Yukawa, or screened Coulomb potential, is often used as a model for fluids
composed of charged particles immersed in a uniform neutralising background.
We define the pair potential as
\begin{equation}
\phi(r)\,=\,\frac{\epsilon \exp(-\lambda r)} {\lambda r},
\label{eq:yplt}
\end{equation}
where $r$ is the distance between the centres of a pair of particles, $\lambda$
is the inverse decay length (screening parameter) and $\epsilon>0$ characterises
the strength of the potential. A fluid composed of particles interacting via
this potential exhibits behaviour similar to that of
the one component classical plasma (OCP) \cite{baushansen1980,levin2002},
in which $\phi(r)\propto 1/r$. The point Yukawa pair potential also corresponds
to a limiting case of the well established Derjaguin, Landau, Verwey and
Overbeek (DLVO) model for aqueous suspensions of charged colloidal particles
\cite{HansenLowen99,hynninen:dijkstra2003,kleinetalJPCM2002}, where the
hard-core part of the colloid-colloid effective pair potential is neglected.

The Yukawa pair potential is also widely employed in theoretical studies of the
so-called dusty plasmas.
These are multicomponent plasmas consisting of charged (dust)
particles, electrons and ions, as well as neutral atoms or molecules, which are
found in a variety of environments, from the interstellar
medium to plasma etching processes. Depending
on their size, the dust grains can attain a large negative charge of
1000--10000$e$ for particles of size 1--10$\mu m$ \cite{piel:melzer2002}; the
charge is generally negative and is determined by the balance of the absorbed
electron and ion fluxes. Since the dust component of the plasma can be
videoed and tracked directly, dusty
plasmas provide a valuable system for studying both equilibrium phenomena and
collective processes such as transport in a fluid
\cite{piel:melzer2002,Valuina}. In modelling the dusty plasma, the
effective potential between two dust particles of charge $Q$ is taken to be
\cite{rosenfeldPRE1994,farouki:hamaguchi1994}
\begin{equation}
\phi(r)\,=\, \frac {Q^2} {4\pi \epsilon_0 r} \exp(-k_Dr),
\label{eq:ypgk}
\end{equation}
where $k_D^{-1}$ is the Debye length of the background plasma. The
thermodynamics of the system can then be characterised by the dimensionless
parameters
\begin{equation}
\kappa\,=\, k_D a
\hspace{4mm} {\rm and} \hspace{4mm}
\Gamma\,=\, \frac {Q^2} {4\pi \epsilon_0 a k_B T},
\label{eq:gk}
\end{equation}
where $a \equiv (3/(4\pi \rho))^{1/3}$ is the Wigner-Seitz radius, i.e.\ the
mean interparticle distance, which is determined by the average fluid number
density $\rho$. $\Gamma$ is
the coupling (plasma) parameter \cite{farouki:hamaguchi1994}. Note that unlike
the OCP, whose properties depend solely on $\Gamma$, {\em two} parameters are
required in this case.

\begin{figure}
\includegraphics[width=8cm]{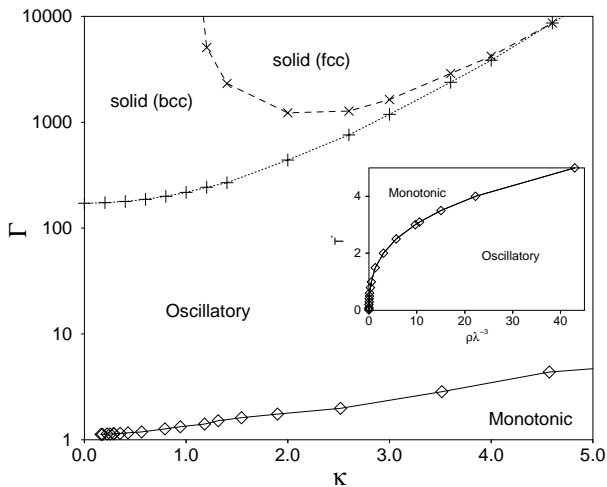}
\caption{\label{fig:phased_full}
Crossover line (diamonds joined by a solid line) separating the region of the
$(\kappa,\Gamma)$ plane where the asymptotic decay of $h(r)$ is damped
oscillatory from that where it is monotonic. In the OCP, corresponding to
$\kappa=0$, crossover occurs at $\Gamma_K=1.21$ \cite{LeoDC}.
We also display the fluid-solid $(+)$ and solid-solid $(\times)$ phase
boundaries given in Ref.\ \cite{HamaguchietalPRE1997}. In the inset we display
the crossover line in the $(\rho, T)$ plane. Note that the freezing transition
present at low $T^* \equiv k_BT/\epsilon$ is not visible on this scale.}
\end{figure}

Because of its relevance for colloidal fluids and the growing importance of
studies of dusty plasmas, not to mention the appealing mathematical properties
of the Yukawa potential \cite{rowlinsonphysicaA1989}, much is established for
this model
fluid. The phase diagram has been determined in a number of simulation studies
\cite{kremeretalPRL1986,robbinsetalJCP1988, rosenberg:thirumalaiPRA1987,
meijer:frenkelJCP1991, HamaguchietalJCP1996, HamaguchietalPRE1997}. At small
values of $\Gamma$ the liquid
state equilibrium structure is well-approximated by the hypernetted-chain
(HNC) closure to the Ornstein-Zernike (OZ) equation \cite{daughtonetalPRE2000,
rosenfeld:ashcroftPRA1979, HanMcD}, as one would expect from studies of the
OCP \cite{baushansen1980}. For larger values of $\Gamma$, a modified HNC based
upon a re-scaling of the bridge function for the OCP \cite{iyetomietalPRA1992}
yields results
indistinguishable from Monte-Carlo (MC) simulation data
\cite{daughtonetalPRE2000}. At small
values of $\kappa$ the Yukawa fluid freezes into a bcc crystal upon increasing
$\Gamma$, as in the OCP. For sufficiently large $\kappa$ increasing $\Gamma$ can
lead to freezing directly into a fcc crystal. A portion of the phase diagram
from Ref.\ \cite{HamaguchietalPRE1997} is shown in Fig.\ \ref{fig:phased_full}
\cite{footnote1}.

In the present work we analyse the asymptotic decay of the pair correlations in
the uniform
fluid and we find that at couplings $\Gamma$ below freezing, there is a
crossover in the form of the asymptotic decay, $r \rightarrow \infty$, of
$h(r)$, the total correlation function, from monotonic to damped
oscillatory, that is similar to the crossover found near $\Gamma_K=1.12$
in the HNC treatment of the OCP \cite{LeoDC}. We map-out the crossover line in
the $(\kappa,\Gamma)$ phase diagram -- see Fig.\ \ref{fig:phased_full}.

\begin{figure}
\includegraphics[height=4.4cm, width=8cm]{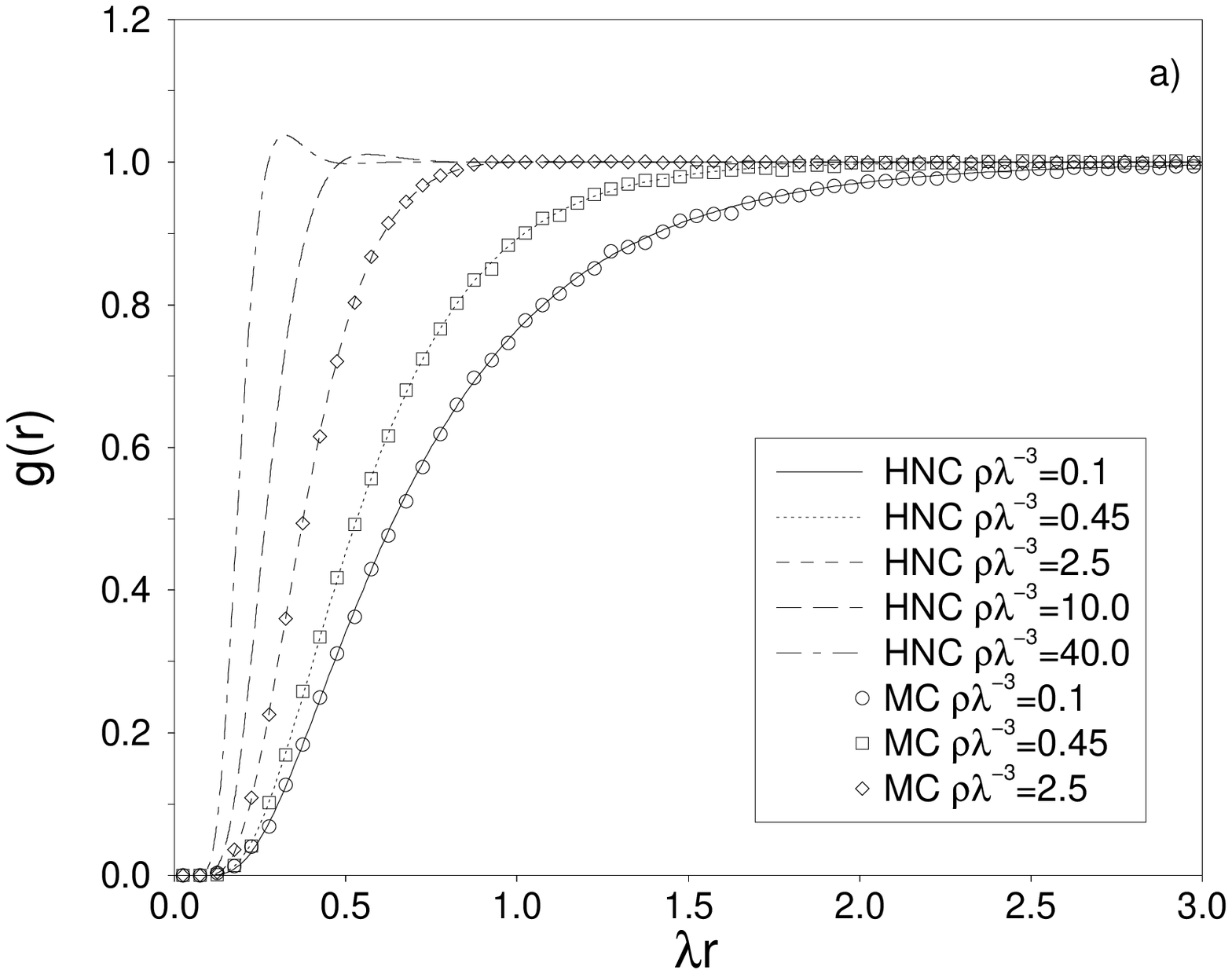}
\includegraphics[height=4.4cm, width=8cm]{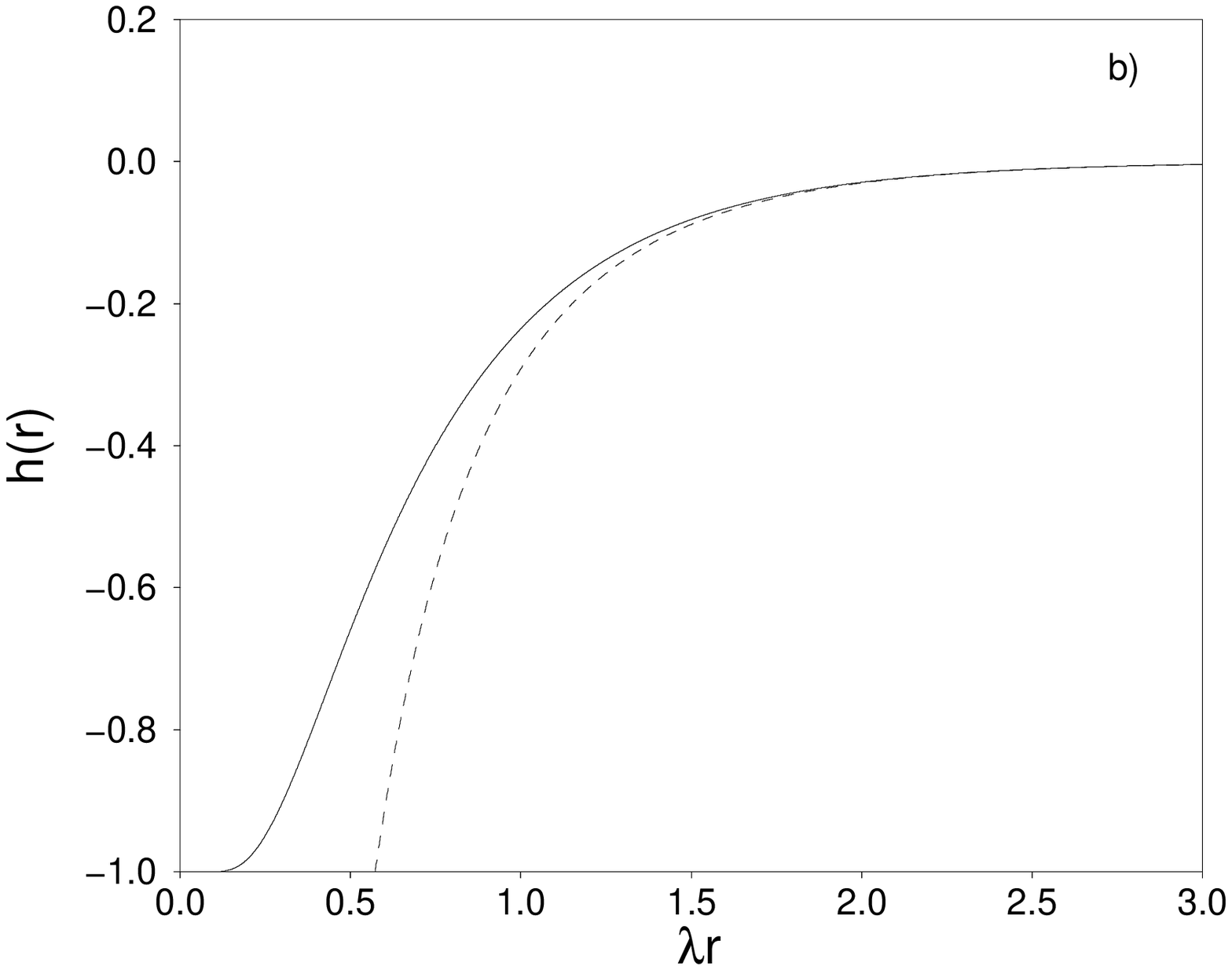}
\includegraphics[height=4.4cm, width=8cm]{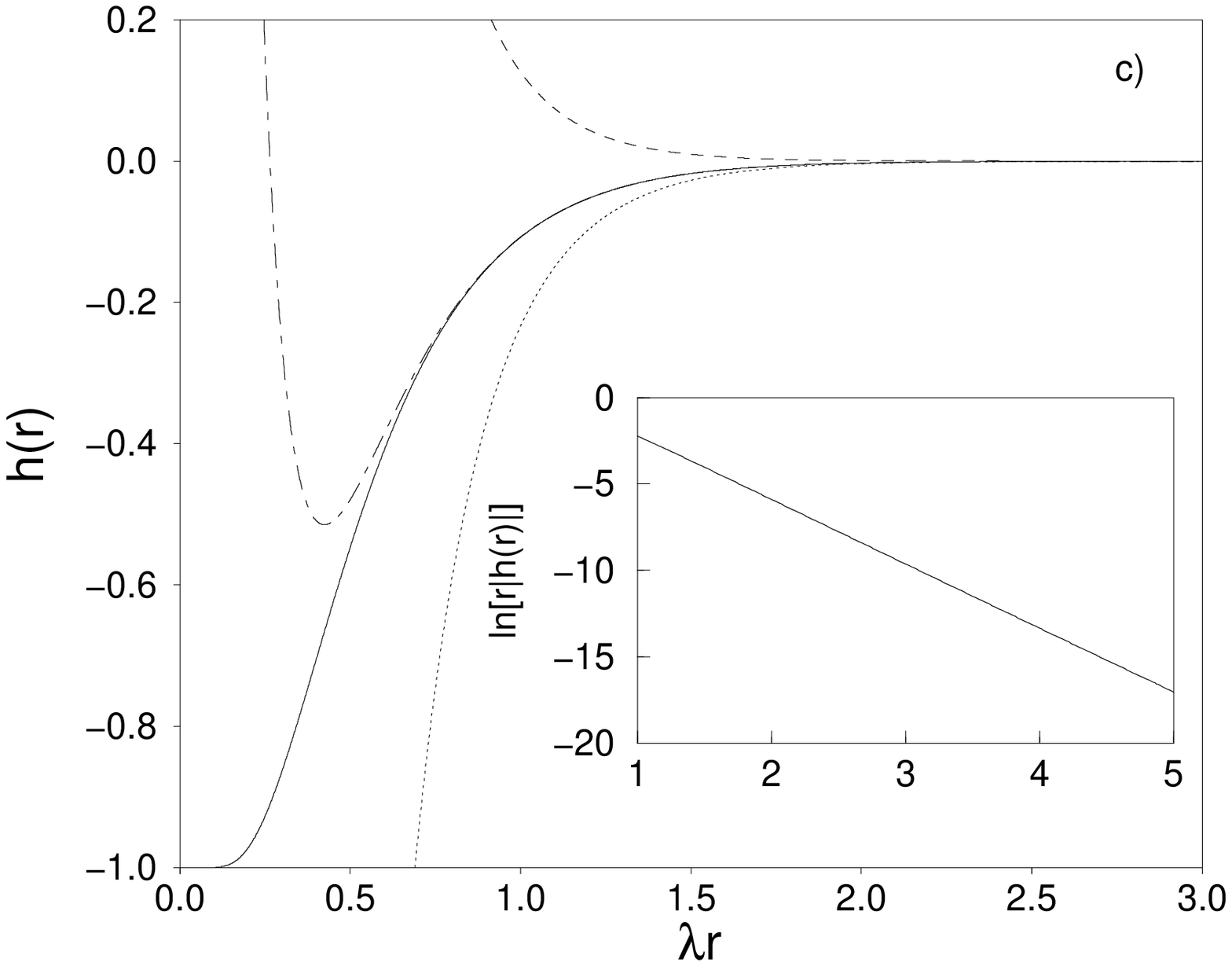}
\includegraphics[height=4.4cm, width=8cm]{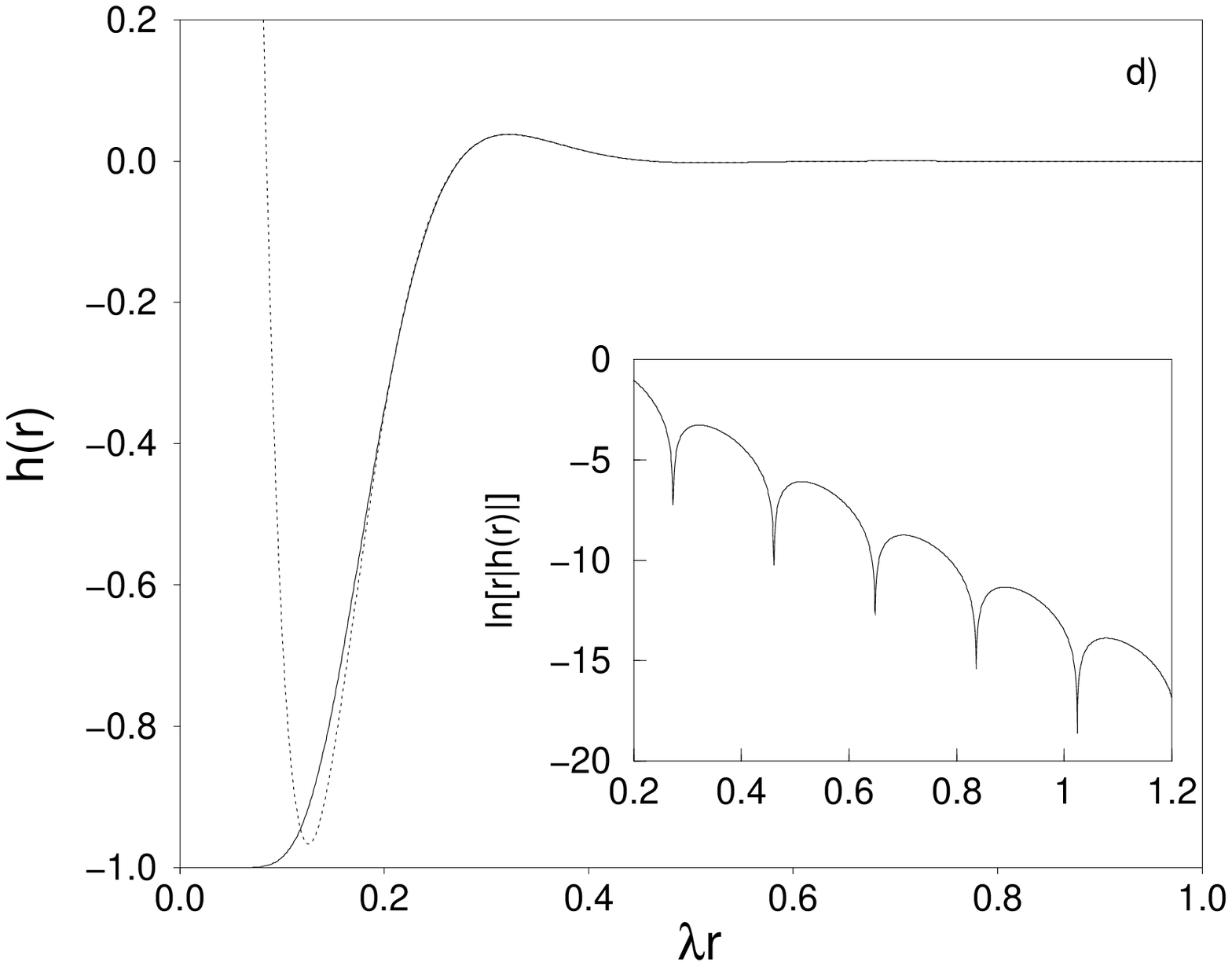}
\caption{\label{fig:gr_set}
Pair correlation functions for a reduced temperature $T^*=1$.
a) $g(r)$ calculated for a range of densities; the lines denote HNC results and
the symbols MC results (see legend). Crossover occurs at
$\rho \lambda^{-3}\simeq0.47$ for this temperature.
For densities larger than this the decay of $g(r)$ is damped oscillatory. b)
Comparison between the full HNC result for $h(r)$ (solid line) and the
contribution from the leading order single
imaginary pole (dashed line) for $\rho \lambda^{-3}=0.1$. c)
The full HNC result for $h(r)$ (solid line) and the
contributions from the two
leading order purely imaginary poles (dotted and dashed lines) and their sum
(dot-dashed line), for $\rho \lambda^{-3}=0.45$. d) The full
HNC result for $h(r)$ (solid line)
and the contribution from the leading order conjugate pair of complex
poles (dotted line), giving damped oscillatory decay, for $\rho
\lambda^{-3}=40$. The insets in c) and d) show $\ln [r|h(r)|]$ versus
$\lambda r$.}
\end{figure}

Our starting point is the OZ equation \cite{HanMcD}, which relates $h(r)$ to
$c(r)$, the pair direct correlation function.
In Fourier space the OZ equation can be written as:
\begin{equation}
\hat{h}(q)\,=\,\frac{\hat{c}(q)} {1-\rho \hat{c}(q)}.
\label{eq:oz2}
\end{equation}
Here $\hat{f}(q)$ denotes the 3-dimensional Fourier transform of the
spherically symmetric
function $f(r)$. We choose to implement the HNC closure which sets the bridge
function $B(r)=0$. For the present Yukawa fluid,
particularly at the densities in the neighbourhood of the crossover line, the
HNC accounts very well for the structure of the uniform fluid, yielding pair
correlation functions almost
indistinguishable from simulation results. Thus we shall use the HNC results
to determine the location of the crossover line. First in Fig.\ \ref{fig:gr_set}
a) we display some typical results for $g(r)=1+h(r)$. These are for a fixed
temperature $T^* \equiv k_BT/\epsilon=1$ and
for the densities $\rho \lambda^{-3}=0.1$, 0.45, 2.5, 10 and 40. At this
temperature, we shall find that the crossover occurs for $\rho \lambda^{-3}
\simeq0.47$. We also display some constant NVT MC results for the first three of
these densities; for $\rho \lambda^{-3}=0.1$ and 0.45 we used 300 particles
with 25000 (plus 10000 equilibration) trial moves per particle. The maximum
displacement of the particles was chosen so that approximately 50\% of attempted
moves were accepted. For the MC simulations
at $\rho \lambda^{-3}=2.5$ we used 2000 particles. We see as the density
is increased the asymptotic decay switches from monotonic to damped
oscillatory. This crossover represents the evolution of the system from a
weakly coupled state where the
particles do not order strongly, to a state where they are more closely
packed and the correlations become more hard-sphere like, although for all the
densities displayed, the amplitude of the (oscillatory) structure in $g(r)$
remains quite small.

The asymptotic decay of $h(r)$ can be obtained \cite{LeoDC,EvansetalJCP1994}
from the OZ equation using the inverse Fourier
transform of Eq.\ (\ref{eq:oz2}):
\begin{equation}
rh(r)\,=\,\frac{1}{4\pi^2i}\int^{\infty} _{-\infty} {\mathrm d} q\,q\,\exp(iqr)\,
\frac{\hat{c}(q)}{1-\rho \hat{c}(q)},
\label{eq:ozint}
\end{equation}
which can be transformed into a semi-circular contour integral in the
upper half of the complex plane. Its value is determined
by the poles of $\hat{h}(q)$ enclosed. These occur at
$q_n=\pm \alpha_1 + i\alpha_0$, where $q_n$ is the solution to the
equation
\begin{equation}
1-\rho \hat{c}(q_n)\,=\,0.
\label{eq:polecrit}
\end{equation}
As a result, $h(r)$ can be obtained as the sum \cite{LeoDC,EvansetalJCP1994}
\begin{equation}
rh(r)\,=\,\frac{1}{2\pi}\sum _n R_n \exp(iq_nr),
\label{eq:tcfsum}
\end{equation}
where $q_n$ is the {\it n}th pole and $R_n$ is the residue of
$q\hat{c}(q)/(1-\rho\hat{c}(q))$ at $q_n$. Clearly the asymptotic behaviour
of $h(r)$ is
determined by the pole(s) with the smallest imaginary part, $\alpha_0$. If
this pole is purely imaginary, $q_n=i\alpha_0$, then $rh(r)\sim
A \exp(-\alpha_0r)$, for $r\rightarrow\infty$, where $A$ is a (real) amplitude
\cite{EvansetalJCP1994}. Alternatively, if the conjugate pair $q_n=\pm\alpha_1+
i\tilde{\alpha}_0$ has the smallest imaginary part then
$rh(r)\sim\tilde{A}\exp(-\tilde{\alpha}_0r)\cos(\alpha_1r-\theta)$. The
amplitudes $A$ and $\tilde{A}$ and the phase $\theta$ can be calculated from the
residues \cite{EvansetalJCP1994}. Whether a pure imaginary or complex pole
dominates depends on the thermodynamic state point.

In order to calculate the poles, we use the separation method introduced in
Ref.\ \cite{LeoDC}. Owing to the particular form of the decay of
the Yukawa pair potential, the asymptotic behaviour of $c(r)$ must be treated
separately so as to ensure the convergence of the integrals which determine the
poles \cite{LeoDC}. The asymptotic decay, $r \rightarrow \infty$, of
the direct correlation function is given by $c(r) \sim -\beta\phi(r)$,
which for the Yukawa potential implies $c(r) \sim -\exp(-\lambda
r)/(T^*\lambda r)$. It is convenient to define a short-ranged direct correlation
function $c^{sr}(r)$ by subtracting the long-ranged Yukawa decay.
The Fourier transform of $c(r)$ is then
\begin{equation}
\hat{c}(q) \equiv \hat{c}^{sr}(q) \,-\,
\frac{4\pi}{\lambda T^*}\frac{1}{(q^2+\lambda^2)}.
\label{eq:csrf}
\end{equation}
Making this division, we follow Ref.\ \cite{LeoDC} and calculate the
poles by separating Eq.\ (\ref{eq:polecrit}) into its real and imaginary parts
and solving numerically using a Newton-Raphson procedure. However, in general
the integrals
involved converge only for complex $q$ such that $\Im[q]<2\alpha_0$, where
$\alpha_0$ is the imaginary part of the leading order pole, i.e.\ that with
the smallest value of $\alpha_0$. In practice, the other poles generally lie
outside this region of convergence and only the leading order pole
can be determined. We also use this separation of $c(r)$ to calculate the
amplitude and phase of $h(r)$ from the residues of the poles, assuming these to
be simple \cite{LeoDC}.

Using our HNC results for $c(r)$ we were able to calculate the contributions to
$h(r)$ from the leading order pole(s) for various points in the phase diagram.
In Fig.\ \ref{fig:gr_set} b) we display the HNC result for $h(r)$ at the state
point $\rho \lambda^{-3}=0.1$ and $T^*=1$, together with the contribution from
the leading order (purely imaginary) pole. This state point lies on the
monotonic side of the crossover line. In Fig.\ \ref{fig:gr_set} c) we display
the HNC result for $h(r)$ at
$\rho \lambda^{-3}=0.45$ and $T^*=1$ (near the crossover line, still on the
monotonic side), together with the contributions from
the two leading order (purely imaginary) poles.
In Fig.\ \ref{fig:gr_set} d) we display the HNC result for $h(r)$, for the state
point $\rho \lambda^{-3}=40$ and $T^*=1$, which exhibits
damped oscillatory asymptotic decay, together with the contribution from
the leading order conjugate pair of complex poles. Note that the contribution
from this pair of poles approximates accurately the full $h(r)$ for
$\lambda r > 0.2$.

\begin{figure}
\includegraphics[width=8cm]{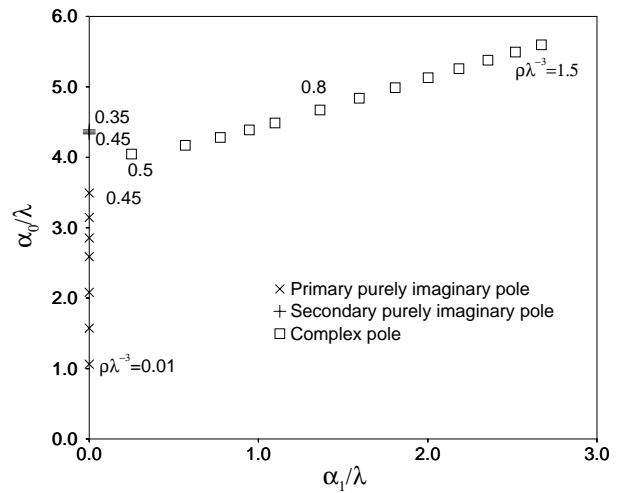}
\caption{\label{fig:poles_evo}
The leading order poles (smallest imaginary part, $\alpha_0$) along the
isotherm $T^*=1$ for increasing $\rho \lambda^{-3}$. Only poles with positive
$\alpha_1$ are shown; complex poles occur in conjugate pairs. At low densities
the leading order pole $(\times)$ is purely imaginary. As the density is
increased a second purely imaginary pole $(+)$ descends and, at
a density $\rho \lambda^{-3} \simeq 0.47$, the two purely imaginary poles
coalesce to form a pair of complex poles $(\Box)$. At this (Kirkwood) point the
asymptotic decay of $h(r)$ crosses over from monotonic to damped oscillatory.}
\end{figure}

In Fig.\ \ref{fig:poles_evo} we display the leading order poles calculated along
the isotherm $T^*=1$. At low densities the pole dominating the
decay of $h(r)$ is purely imaginary. However, as the density
is increased, this pole moves up the imaginary axis, and at a density $\rho
\lambda^{-3} \simeq 0.47$ it meets a second (descending) purely imaginary pole.
This second pole could only be determined once it had descended into the region
of convergence. These poles coalesce and form a conjugate pair of complex poles
which then move away from the imaginary axis as the density is increased
further. This coalescence of two purely imaginary poles to form a pair of
complex poles as one moves along a path of increasing density in the phase
diagram, results in a crossover in the asymptotic decay of
$h(r)$ from monotonic to damped oscillatory. The mechanism of poles coalescing
and moving off the imaginary axis is the same as that found in the OCP near
$\Gamma_K=1.12$ \cite{LeoDC}. It is equivalent to that found in studies of
charge correlations in the restricted primitive model (RPM) of binary ionic
fluids \cite{LeoDC:EvansMolP94} and in screened versions of the RPM
\cite{LeoDC:EvansMolP97}. Kirkwood \cite{Kirkwood} was the first to describe the
mechanism, so the line in the phase diagram at which crossover occurs
could be termed the Kirkwood line following earlier terminology
\cite{LeoDC:EvansMolP94,LeoDC:EvansMolP97,LeoDC}.

In Fig.\ \ref{fig:phased_full} we display the location of the crossover
(Kirkwood) line in the phase diagram. This line (see inset) was determined by
finding the density at which the imaginary poles coalesce for a series of
isotherms, i.e.\ fixed $T^*$. By converting from the variables $(\rho
\lambda^{-3},T^*)$ to $(\kappa, \Gamma)$ the crossover line shown in the main
figure was obtained. We emphasise that this line lies far below the fluid-solid
transition line, as might have been expected from the observation that in the
OCP freezing is known to occur for $\Gamma \simeq
172$, whereas crossover from monotonic to oscillatory decay of correlations
occurs near $\Gamma_K=1.12$. For the values of $\kappa$ considered here, $0
\leq \kappa \lesssim 5$, crossover occurs for $\Gamma$ in the range $1 \lesssim
\Gamma \lesssim 5$, i.e.\ the coupling parameter is rather weak and we expect the
HNC to perform rather well; see also Fig.\ \ref{fig:gr_set} a) for the
comparison with MC. Of course, one could attempt to improve upon the HNC by
incorporating an approximate bridge function $B(r)$, along the lines of Refs.\
\cite{daughtonetalPRE2000,rosenfeld:ashcroftPRA1979}. However, we do not expect
significant changes in the location of the crossover line; the poles should not
be sensitive to details of the particular closure \cite{LeoDC}. One could also
attempt to extract the poles and determine the crossover line using accurate
simulation data for $g(r)$ following the method applied in Ref.
\cite{DijkstraEvansJCP2000} for a truncated Lennard-Jones fluid.

In summary, we have shown that the form of the asymptotic decay of the total
correlation function $h(r)$ in a Yukawa fluid crosses over from monotonic at
small coupling parameter $\Gamma$ to damped oscillatory at larger values via the
same mechanism as in the OCP.
We find that leading-order asymptotics provide as accurate a description of pair
correlations at intermediate range in the Yukawa fluid as they do for other
model fluids \cite{LeoDC, EvansetalJCP1994, LeoDC:EvansMolP94,
LeoDC:EvansMolP97, DijkstraEvansJCP2000}. This observation might prove useful in
further applications of the Yukawa model to dusty plasmas.

\section*{Acknowledgements}
AJA gratefully acknowledges the support of EPSRC under grant number
GR/S28631/01. RE thanks M. Dijkstra and R. Roth for instructive discussions.

\end{document}